\documentclass[aps,pra,showpacs,twocolumn,superscriptaddress]{revtex4-1}
\usepackage{amsmath}
\usepackage{amssymb}
\usepackage{graphicx}
\usepackage{epstopdf}
\usepackage{times,txfonts}
\usepackage{easybmat}
\usepackage{mathtools}
\usepackage{subfigure}

\usepackage[bookmarks=false]{hyperref}
\hypersetup{colorlinks=true,citecolor=blue,linkcolor=blue,urlcolor=blue,pdfstartview=FitH,bookmarksopen=true}
\usepackage{color,soul}

\begin{document}
\preprint{APS/123-QED}

\title{Proof-of-principle experimental demonstration of quantum gate verification}

\author{Maolin Luo}
\author{Xiaoqian Zhang}
\email{zhangxq67@mail.sysu.edu.cn}
\author{Xiaoqi Zhou}
\email{zhouxq8@mail.sysu.edu.cn}
\affiliation{School of Physics and State Key Laboratory of Optoelectronic Materials and Technologies, Sun Yat-sen University, Guangzhou 510000, China}

\date{\today}

\begin{abstract}
   To employ a quantum  device, the performance of the quantum gates in the device needs to be evaluated first. Since the dimensionality of a quantum gate grows exponentially with the number of qubits, evaluating the performance of a quantum gate is a challenging task. Recently, a scheme called quantum gate verification (QGV) has been proposed, which can verifies quantum gates with near-optimal efficiency. In this paper, we implement a proof-of-principle optical experiment to demonstrate this QGV scheme. We show that for a single-qubit quantum gate, only $\sim300$ samples are needed to confirm the fidelity of the quantum gate to be at least $97\%$ with a $99\%$ confidence level using the QGV method, whereas, at least $\sim3000$ samples are needed to achieve the same result using the standard quantum process tomography method. The QGV method validated by this paper has the potential to be widely used for the evaluation of quantum devices in various quantum information applications.
\end{abstract}

\maketitle

Quantum information technology can greatly improve the speed of computation, the security of communication, and the precision of measurement. To make good use of quantum information technology, it is first necessary to characterize quantum devices to evaluate their performance. The standard method for characterizing quantum devices is quantum process tomography (QPT) \cite{1Deville,2Lo,3Teo,4Hou,5Bouchard,6Gazit,7Thinh}, which allows complete reconstruction of the quantum process of the device, but its resource overhead grows exponentially with the size of the system, making it impractical when the system is large. However, for most applications, the complete information of the quantum device is not needed, but only the fidelity of the evaluated device compared to a perfect device. For this reason, methods such as direct fidelity estimation \cite{9Flammia} and randomized benchmarking \cite{11Knill,12Magesan,13Wallman,14Harper,15Onorati,16Helsen} have been proposed to estimate the fidelity of quantum gates. Although these methods improve the efficiency of the verification of quantum gates, they require a very large number of experimental settings and are not optimal in terms of the scaling behavior between the predicted upper limit of the infidelity $\epsilon$ and the number of samples N.

Recently, a quantum gate verification (QGV) \cite{32Zhu,33Liu,34Zeng} scheme developed from the quantum state verification (QSV) \cite{17Pallister,18Zhu,19Zhu,20Wang,21Yu,22Li,23Liu,24Li,25Zhu,26Zhu,27Dangniam,28Liu,29Zhang,30Jiang,31Zhang} scheme has been proposed. This scheme allows gate verification of a variety of quantum gates with near-optimal efficiency using only local state preparation and local measurements. In this paper, we implement a proof-of-principle optical experiment to demonstrate this QGV scheme. We randomly selected two arbitrary single-qubit  gates and performed QGV on them. It has been show that only $\sim300$ samples are needed to confirm the fidelity of the quantum gate to be at least $97\%$ with a $99\%$ confidence level, whereas, at least, $\sim3000$ samples are needed to achieve the same result using the QPT method. In addition, the QPT method requires 18 experimental settings compared to the QGV method which only requires 6 experimental settings. Our results demonstrate that quantum gates can be efficiently verified using the QGV method.

\emph{Theoretical framework for quantum-gate verification.}---We first briefly review how to transform a QGV problem into a QSV problem \cite{32Zhu,33Liu,34Zeng} .
As shown in Fig. 1(a), k qubits of a 2k-qubit maximally-entangled bipartite state $\rho$ are passed through a quantum process $\Lambda_{ideal}$ thereby obtaining an output state $\rho_{ideal}$.
If no other quantum process except $\Lambda_{ideal}$ can convert $\rho$ to $\rho_{ideal}$ in this way,  then the quantum process $\Lambda_{ideal}$ corresponds exactly to the quantum state $\rho_{ideal}$.
Therefore, the problem of verifying whether a quantum process $\Lambda_{device}$ is equal to $\Lambda_{ideal}$ is transformed into the problem of verifying whether the output state $\rho_{device}$ obtained from the quantum state $\rho$ through the quantum process $\Lambda_{device}$ is equal to $\rho_{ideal}$.

\begin{figure}[!htp]
  \centering
  \includegraphics[width=0.4\textwidth]{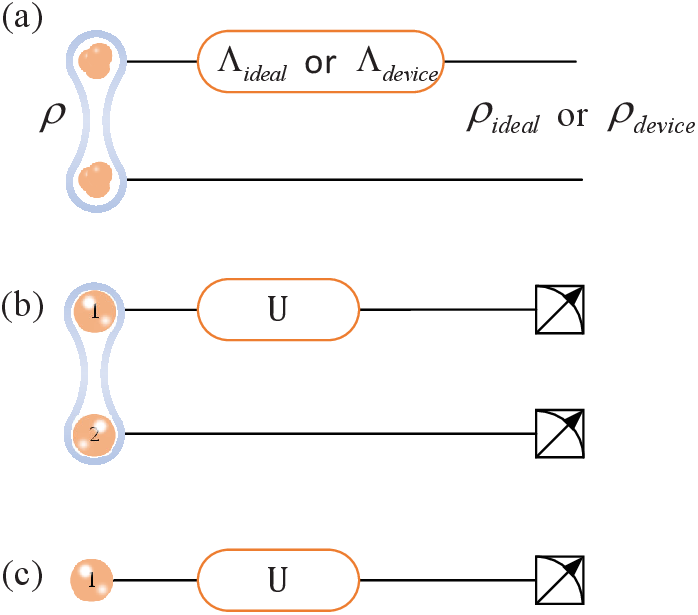}
  \caption{The schematic of quantum gate verification. (a) Transformation of quantum gate verification problem to quantum state verification problem. By using the entangled probing state $\rho$, the problem of whether the quantum process $\Lambda_{device}$ is equal to $\Lambda_{ideal}$ is transformed into the problem of whether the output quantum state $\rho_{device}$ is equal to $\rho_{ideal}$. (b) The quantum gate verification of a single-qubit unitary $U$ using a two-qubit entangled probing state.  (c) The quantum gate verification of a single-qubit unitary $U$ using only single-qubit probing states.}\label{}
\end{figure}

In the following, we take an arbitrary single-qubit gate $U$ as an example to illustrate the method of QGV in detail.
As shown in Fig. 1(b), the two-qubit state $\phi$ is used as the input state, where
\begin{eqnarray}
\phi=\frac{1}{2}(|0\rangle_1|0\rangle_2+|1\rangle_1|1\rangle_2).
\end{eqnarray}
After passing qubit 1 through a quantum process $\Lambda_{device}$, the output two-qubit state $\phi_{\Lambda_{device}}$ would be equal to $\phi_{U}$ if $\Lambda_{device}$ is equal to the $U$ operation, in which
\begin{eqnarray}
\phi_{U}=\frac{1}{2}(U|0\rangle_1\otimes |0\rangle_2+U|1\rangle_1\otimes |1\rangle_2).
\end{eqnarray}
The efficient strategy for verifying whether $\phi_{\Lambda_{device}}$ is equal to $\phi_{U}$ can be defined by the operator $\Omega_{\phi_{U}}=\frac{1}{3}(P_{(UXU^\dagger)_1X_2}^+ +P_{(UYU^\dagger)_1Y_2}^- +P_{(UZU^\dagger)_1Z_2}^+)$. It can be realized by randomly choosing one of the three bases $(UXU^\dagger)_1X_2$, $(UYU^\dagger)_1Y_2$ and $(UZU^\dagger)_1Z_2$ for measurement. When the measurement basis is $(UXU^\dagger)_1X_2$ or $(UZU^\dagger)_1Z_2$, the measurement result $+1$ is taken as a pass, and when the measurement basis is $(UYU^\dagger)_1Y_2$, the measurement result $-1$ is taken as a pass. Suppose we have N copies of $\phi_{\Lambda_{device}}$, the verification is passed only when all N measurement outcomes are passed, at which point it can be claimed that the fidelity of $\phi_{\Lambda_{device}}$ with respect to $\phi_{U}$ is at least $1-\epsilon$, with $1-\delta$ confidence, where
\begin{eqnarray}
\delta\leqslant e^{-\epsilon N\nu},
\end{eqnarray}
in which $\nu$ is the difference between the largest and the second largest eigenvalues of
$\Omega_{\phi_{U}}$.

The original scheme above requires all measurement outcomes to be passed, which is difficult to achieve in practice, so we will use a modified version of the scheme \cite{29Zhang} that allows for failed outcomes. If among N measurement outcomes, M are passed, then when $\epsilon\geqslant \frac{1-M/N}{\nu}$, it can be claimed that the fidelity is at least $1-\epsilon$, with $1-\delta$ confidence, where
\begin{eqnarray}
\delta\leqslant e^{-D(\frac{M}{N}||1-\epsilon\nu)N},
\end{eqnarray}
with
\[
D(x||y)=xlog_2\frac{x}{y}+(1-x)log_2\frac{1-x}{1-y}.
\]

By analyzing the above scheme, it can be found that the measurements on qubit 2 can occur before qubit 1 pass through the quantum gate, which means the scheme of preparing a two-qubit entangled state as the input state can be transformed into a scheme of preparing some single-qubit states as the input state. Taking the $(UZU^\dagger)_1Z_2$ measurement setting as an example, measuring qubits 1 and 2 of state $|\phi\rangle$ first is equivalent to preparing a single-qubit state $|0\rangle$ or $|1\rangle$ randomly with equal probability, letting it pass through the quantum gate, and then performing a $UZU^\dagger$ measurement on it. Therefore, the verification scheme for the single-qubit gate becomes --- prepare one of the following six single-qubit quantum states $|0\rangle$, $|1\rangle$, $|+\rangle$, $|-\rangle$, $|+i\rangle$ and $|-i\rangle$ randomly with equal probability and let it pass through the quantum gate and then measure it at a certain basis, where $|\pm\rangle=\frac{1}{\sqrt{2}}(|0\rangle\pm|1\rangle)$ and $|\pm i\rangle=\frac{1}{\sqrt{2}}(|0\rangle\pm i|1\rangle)$.
If the initial state is $|0\rangle$ ($|+\rangle$ / $|-i\rangle$), then it is measured at $UZU^\dagger$ ($UXU^\dagger$/$UYU^\dagger$) basis and the verification passes with outcome +1; if the initial state is $|1\rangle$ ($|-\rangle$/$|+i\rangle$), then it is measured at $UZU^\dagger$ ($UXU^\dagger$ / $UYU^\dagger$) basis and the verification passes with outcome -1.
\begin{figure}[!htp]
  \flushleft
  \includegraphics[width=0.45\textwidth]{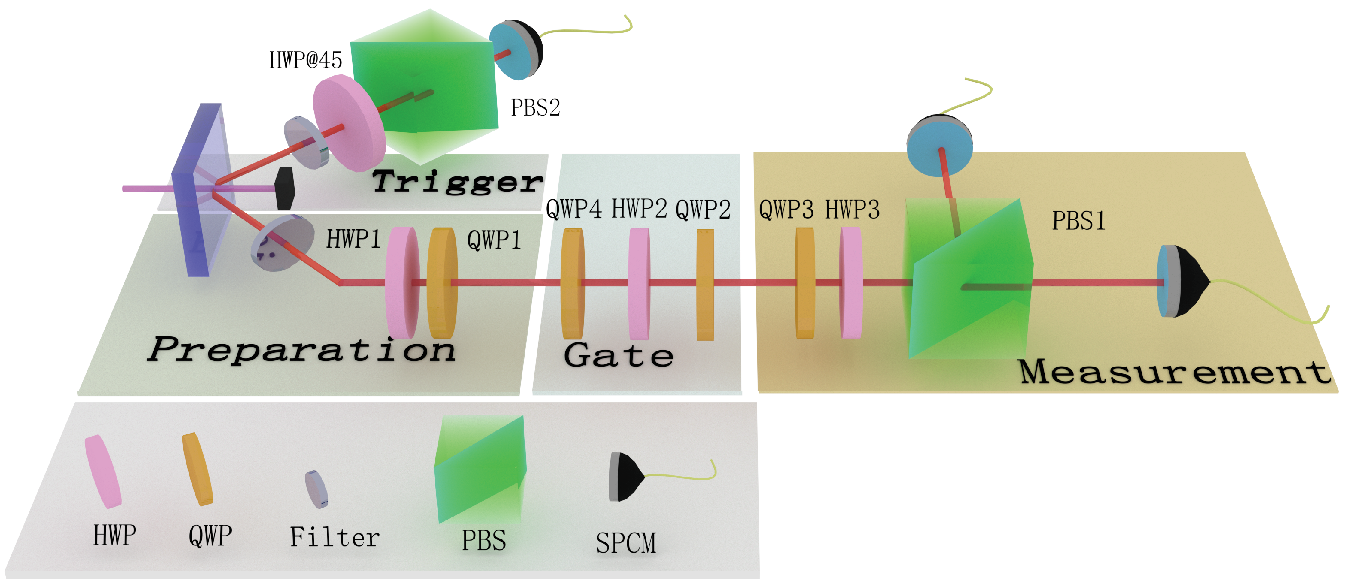}
  \caption{Experimental setup. A continuous ultraviolet laser is focused on a type-II $\beta$-barium borate (BBO) crystal and produces a photon pair. After the triggering of photon 2, photon 1 is prepared at the desired state by tuning half-wave plate 1 (HWP1) and quarter-wave plate 1 (QWP1). After passing through the single-qubit gate, which is realized by QWP4, HWP2, and QWP2, photon 1 is measured by the device [QWP3, HWP3, polarization beam splitter 1 (PBS1), and the two single-photon counting modules (SPCMs)] at the desired basis.}\label{}
\end{figure}

\emph{Experimental setup. }---The experimental setup to implement the single-qubit gate verification is shown in Fig. 2. A 50 mW ultraviolet laser with a central wavelength of 405 nm is focused on a type-II BBO crystal to generate a photon pair $|H\rangle_1|V\rangle_2$, where H and V denote horizontal and vertical polarizations respectively. After the triggering of photon 2, photon 1 is prepared at the desired state via the HWP1 and the QWP1 as the input state for the verification of the quantum gate, which is realized by QWP4, HWP2 and QWP2 (each waveplate has been precisely calibrated before the experiment, and the deviation angle of the optical axis is within $0.5^0$).
After passing through the quantum gate, the output quantum state is measured by the device (QWP3, HWP3, the PBS1 and the SPCMs) at the desired measurement basis (see Appendix A).

\emph{Results. }---Our experiments follow strictly the non-trace-preserving prepare-and-measure scheme of the Ref. \cite{33Liu}.
We choose three general single-qubit gates $U_a$, $U_b$ and $U_c$ for demonstrating the scheme of QGV, where
\begin{eqnarray}
\begin{array}{l}
\displaystyle U_a=\left(
 \begin{array}{cc}
     -0.0360 + 0.3672i      & -0.5460 - 0.7521i\\
     -0.8446 - 0.3880i     &-0.3530 + 0.1073i\\
 \end{array}
 \right),\\
 \\
\displaystyle U_b=\left(
 \begin{array}{cc}
     0.1641 + 0.9256i   &  0.3158 - 0.1289i \\
     -0.3246 - 0.1050i  & 0.0945 - 0.9353i \\
 \end{array}
 \right),\\
 \\
 \displaystyle U_c=\left(
 \begin{array}{cc}
     -0.8634 + 0.3324i     & 0.0169 + 0.3793i \\
     0.0591 - 0.3750i      & 0.8209 + 0.4267i \\
 \end{array}
 \right).
\end{array}
\end{eqnarray}
The matrices of the three single-qubit gates $U_a$, $U_b$ and $U_c$ are generated randomly by calling the function RandomUnitary \cite{John}, and their physical implementation is achieved by using two QWPs and one HWP as shown in Fig.~2.

\begin{figure*}[!htp]
  \flushleft
  \includegraphics[width=0.72\textwidth]{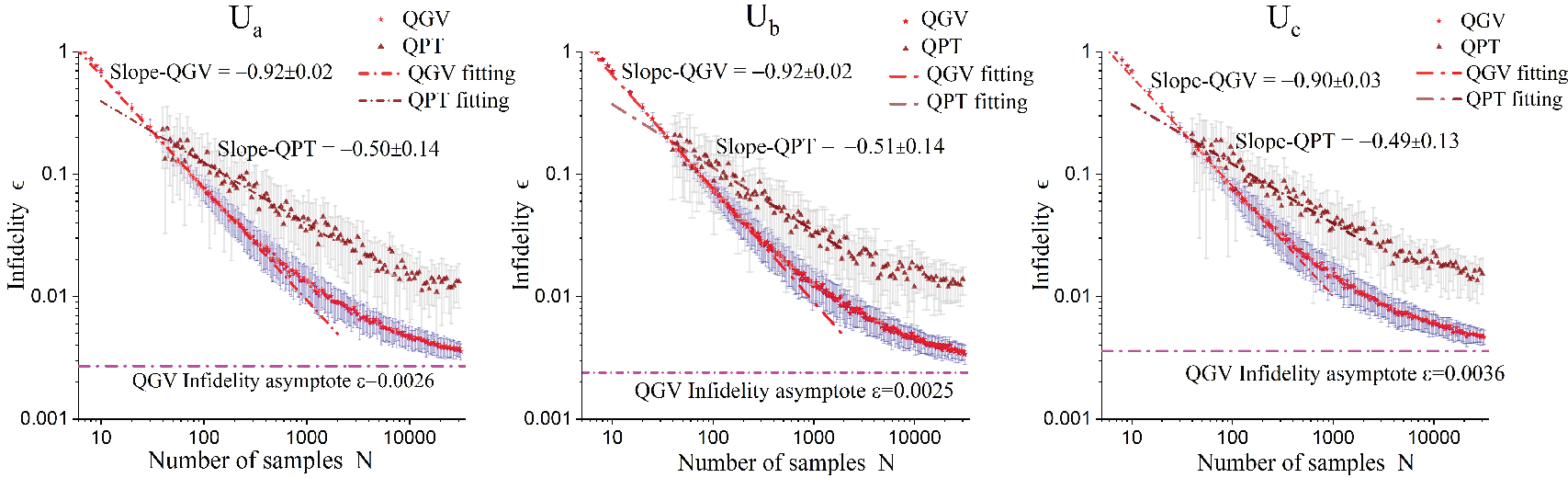}
  \caption{Experimental results of QGV for $U_a$, $U_b$ and $U_c$. The red five-pointed stars correspond to QGV and the brown triangles correspond to QPT. The red and brown lines are the fitted lines to the QGV and QPT data, respectively. Here the confidence level $1-\delta$ is set to $99\%$. The fitting interval is the interval with sample size N less than 500. The fitting formula is $\epsilon\sim N^r$, where $\epsilon$ denotes the infidelity and $r$ denotes the descent slope of the fitted line. For $U_a$ ($U_b$, $U_c$), whereas the QPT's descent slope is just $-0.50\pm0.14$ ($-0.51\pm0.14$, $-0.49\pm0.13$), QGV's descent slope is $-0.92\pm0.02$ ($-0.92\pm0.02$, $-0.90\pm0.03$), which is quite close to the Heisenberg scaling value $-1$.}
\end{figure*}

To verify these gates, we prepare $|H\rangle$, $|V\rangle$, $|D\rangle$, $|A\rangle$, $|R\rangle$ and $|L\rangle$ randomly with equal probability as the input state, in which $|H/V\rangle$ corresponds to $|0/1\rangle$, $|D/A\rangle=\frac{1}{\sqrt{2}}(|H\rangle\pm|V\rangle)$ and $|R/L\rangle=\frac{1}{\sqrt{2}}(|H\rangle\pm i|V\rangle)$. For $U_a$ ($U_b$, $U_c$), we measure the output state on the measurement basis $\chi^+_{a1}/\chi^-_{a1}$ ($\chi^+_{b1}/\chi^-_{b1}$, $\chi^+_{c1}/\chi^-_{c1}$) for the input states $|H\rangle$ and $|V\rangle$ or $\chi^+_{a2}/\chi^-_{a2}$ ($\chi^+_{b2}/\chi^-_{b2}$, $\chi^+_{c2}/\chi^-_{c2}$) for the input states $|D\rangle$ and $|A\rangle$ or $\chi^+_{a3}/\chi^-_{a3}$
\[
\begin{split}
\chi_{a_1}^+&=\left(
 \begin{array}{c}
     -0.0360+0.3672i \\
     -0.8446-0.3880i \\
 \end{array}
 \right), \ \ \chi_{a_1}^-=\left(
 \begin{array}{c}
    -0.5460-0.7521i\\
    -0.3530+0.1073i\\
 \end{array}
 \right),\\
\chi_{a_2}^+&=\left(
 \begin{array}{c}
     -0.4116 - 0.2722i\\
     -0.8468 - 0.1985i\\
 \end{array}
 \right), \ \ \chi_{a_2}^-=\left(
 \begin{array}{c}
     0.3606 + 0.7915i\\
     -0.3476 - 0.3502i\\
 \end{array}
\right),\\
\chi_{a_3}^+&=\left(
 \begin{array}{c}
     0.5064 - 0.1265i \\
     -0.6731 - 0.5240i\\
 \end{array}
 \right), \ \ \chi_{a_3}^-=\left(
 \begin{array}{c}
     -0.5573+0.6457i\\
     -0.5213-0.0248i\\
 \end{array}
 \right),
\\
\chi_{b_1}^+&=\left(
 \begin{array}{c}
     0.1641 + 0.9256i\\
     -0.3246 - 0.1050i\\
 \end{array}
 \right), \ \ \chi_{b_1}^-=\left(
 \begin{array}{c}
     0.3158 - 0.1289i\\
     0.0945 - 0.9353i\\
 \end{array}
 \right),\\
\chi_{b_2}^+&=\left(
 \begin{array}{c}
     0.3394 + 0.5633i\\
     -0.1627 - 0.7355i\\
 \end{array}
 \right), \ \ \chi_{b_2}^-=\left(
 \begin{array}{c}
     -0.1073 + 0.7456i\\
     -0.2963 + 0.5871i\\
 \end{array}
\right),\\
\chi_{b_3}^+&=\left(
 \begin{array}{c}
      0.2072 + 0.8778i\\
      0.4318 - 0.0074i\\
 \end{array}
 \right), \quad \  \chi_{b_3}^-=\left(
 \begin{array}{c}
     0.0249 + 0.4312i\\
     -0.8908 - 0.1410i\\
 \end{array}
\right),\\
\chi_{c_1}^+&=\left(
 \begin{array}{c}
     -0.8634 + 0.3324i \\
     0.0591 - 0.3750i \\
 \end{array}
 \right), \ \ \chi_{c_1}^-=\left(
 \begin{array}{c}
     0.0169 + 0.3793i\\
     0.8209 + 0.42672i\\
 \end{array}
 \right),\\
\chi_{c_2}^+&=\left(
 \begin{array}{c}
     -0.5985 + 0.5032i\\
     0.6223 + 0.0365i\\
 \end{array}
 \right), \ \ \chi_{c_2}^-=\left(
 \begin{array}{c}
     -0.6225 - 0.0332i \\
     -0.5386 - 0.5669i\\
 \end{array}
\right),\\
\chi_{c_3}^+&=\left(
 \begin{array}{c}
     -0.8787 + 0.2470i \\
     -0.2599 + 0.3153i\\
 \end{array}
 \right), \ \ \chi_{c_3}^-=\left(
 \begin{array}{c}
     -0.3423 + 0.2231i\\
      0.3435 - 0.8456i\\
 \end{array}
 \right).
\end{split}
\]
Here $\chi^+_{a1}$ ($\chi^+_{a2}$, $\chi^+_{a3}$) is the eigenstate of the operator $U_aZU_a^\dagger$ ($U_aXU_a^\dagger$, $U_aYU_a^\dagger$) with eigenvalue +1. $\chi^-_{a1}$ ($\chi^-_{a2}$, $\chi^-_{a3}$) is the eigenstate of the operator $U_aZU_a^\dagger$ ($U_aXU_a^\dagger$, $U_aYU_a^\dagger$) with eigenvalue -1.

In order to show the advantages of the QGV method over the traditional QPT method, in addition to the gate verification of $U_a$, $U_b$ and $U_c$ with QGV, we also evaluated the infidelity of $U_a$, $U_b$ and $U_c$ with the QPT method (see Appendix B), and the relevant experimental results are shown in Fig. 3. For a given confidence level $1-\delta$, the infidelity $\epsilon$ decreases as the number of samples N increases, and the faster it decreases, the more effective the method is. It can be seen that for $U_a$, $U_b$ and $U_c$, QGV allows $\epsilon$ to decrease faster with increasing N compared to QPT (the confidence level $1-\delta$ is set to $99\%$). For $U_a$ ($U_b$, $U_c$), the infidelity can be achieved down to 0.03 using the QGV method with only 299 (290, 335) samples, whereas 3133 (2128, 2655) samples are required to achieve the same level of infidelity using the QPT method. To quantify the rate of decrease of $\epsilon$ with N, the data are fitted in the interval $N<500$ using $\epsilon \sim N^r$ \cite{36de,37Mahler,38Kravtsov}, where $r$ is the descent slope of the fitting line. For $U_a$ ($U_b$, $U_c$), whereas QPT's descent slope is just $-0.50\pm0.14$ ($-0.51\pm0.14$, $-0.49\pm0.13$), QGV's descent slope is $-0.92\pm0.02$ ($-0.92\pm0.02$, $-0.90\pm0.03$), which is quite close to the Heisenberg scaling value $-1$ \cite{29Zhang,30Jiang}.

In Ref. \cite{Pogor}, the scaling for QPT on single-qubit gates is around -0.5, where the method used to process the QPT data is Bayesian mean estimation. The method we use to process the QPT data when performing QPT on single-qubit gates is the maximum-likelihood estimation \cite{Qiang}, and the scaling obtained is also around -0.5 \cite{38}. Here QPT experiments are repeated 15 times and QGV experiments are repeated 50 times in order to calculate the error bar shown in Fig. 3. The cumulative measurement time under each measurement basis in the QGV (QPT) experiment is 3.5 h (35 min). In terms of experimental settings, the QGV method is more efficient compared to the QPT method, where 6 experimental settings are used in the above experiments for QGV, whereas 18 experimental settings are used for QPT.

\begin{figure*}[!htp]
  \flushleft
  \qquad\qquad\includegraphics[width=1\textwidth]{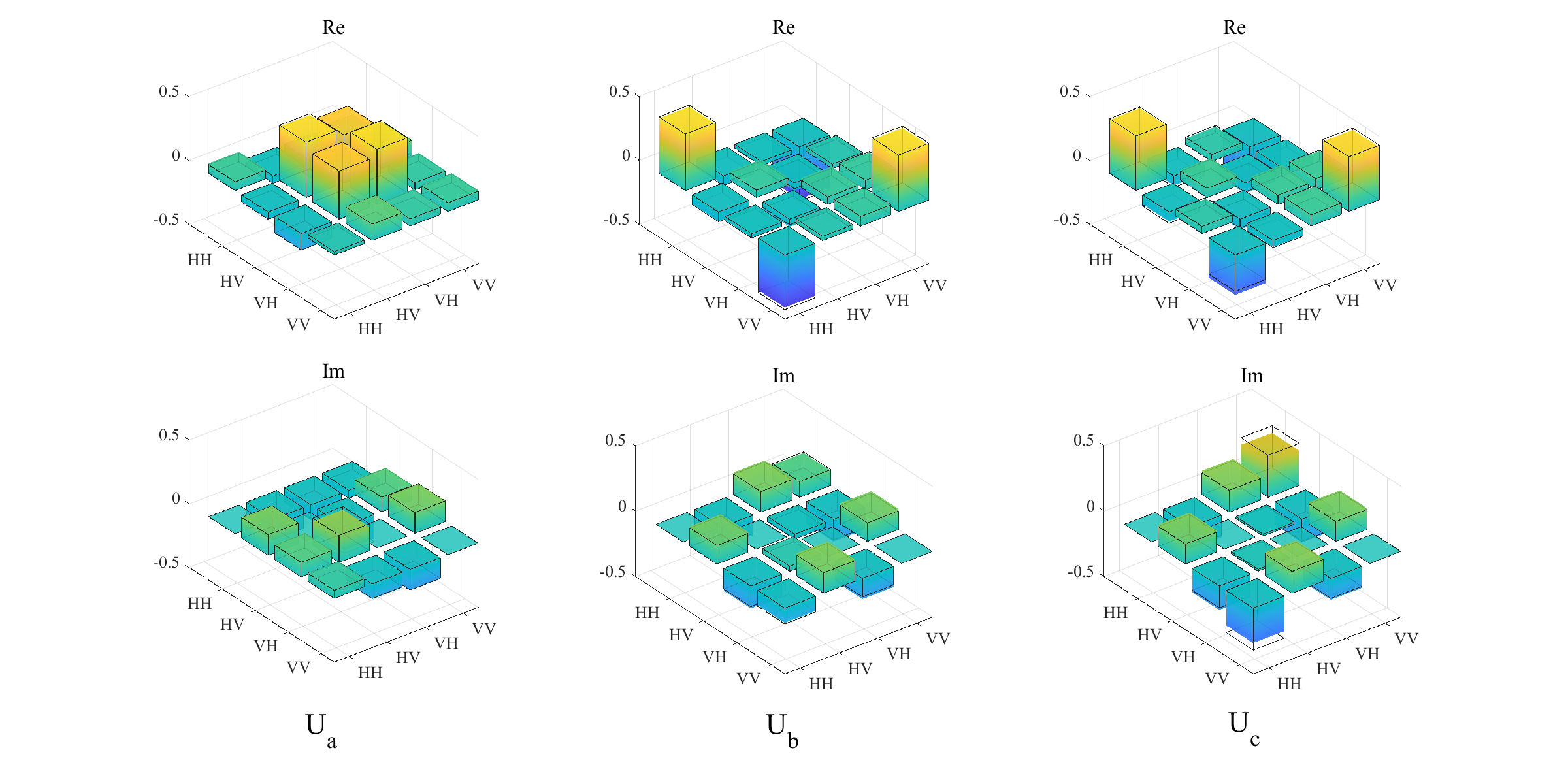}
  \caption{QPT results for $U_a$, $U_b$ and $U_c$. Here ``Re" and ``Im" refer to the real and imaginary parts of the process matrices, respectively. The fidelity of $U_a$ is $0.9902\pm0.0026$, $U_b$ is $0.9891\pm0.0027$, $U_c$ is $0.9872\pm0.0029$. The error bar is calculated by repeating the QPT experiment 100 times.}\label{}
\end{figure*}

In addition to the gate verification of single-qubit gates, we also verify a two-qubit CNOT gate using the QGV method (see Appendix C).
Although QGV does not show an advantage over QPT in terms of the number of samples in this CNOT experiment, QGV still shows a great advantage in terms of the number of experimental settings, with only 16 experimental settings required for QGV, compared to 324 experimental settings for QPT.

\emph{Summary.}---We have experimentally demonstrated the QGV protocol that enables efficient verification of quantum gates. Our experimental results show that the evaluation of quantum gate infidelity can approach the Heisenberg scaling (1/N) which greatly saves the resources needed to evaluate the fidelity of quantum gates compared to the standard QPT method. QGV uses much less measurement bases than QPT, and in practice, switching the measurement basis is time and resource consuming. Even without considering the time and resources spent on switching measurement basis, the gate fidelity value estimated by QGV is closer to the real fidelity value than that estimated by QPT for the same measurement time. The QGV method validated by this paper is expected to be widely used for the evaluation of quantum devices and for various quantum information applications including quantum process discrimination \cite{41Anthony}, quantum channel quantification \cite{42Daz} and quantum entanglement detection \cite{43Ghnea}.\\

This work was supported by the National Key Research and Development Program (2017YFA0305200), the Key Research and Development Program of Guangdong Province of China (2018B030329001 and 2018B030325001), the National Natural Science Foundation of China (Grant No. 61974168). X.-Q. Zhou acknowledges support from the National Young 1000 Talents Plan. X.-Q. Zhang acknowledges support from the National Natural Science Foundation of China (Grant No. 62005321). \\

\section*{Appendix A: Details of the experimental setup for single-qubit gates}

The details for the single-qubit gates experimental setup are listed as follows:\\
(1) The calculated spectral widths of both ordinary-light (o-light) and extraordinary-light (e-light) are 2.47nm and the BBO crystal length is 2mm.\\
(2) The 5-nm filters are used to filter the spontaneous parametric down-conversion light and the coincidence rate is around 4700/s with 5-ns coincidence window. The accidental coincidence rate is 1.305/s. \\
(3) The photon pair generation rate is 82076/s. The probability of producing one pair of photons in one measurement interval (20 ns) is $p=1.64\times 10^{-3}$, and the probability of producing two pairs of photons is $p^2$.\\
(4) The heralding efficiencies of the two channels are $\eta_1=R_c/R_1=23.71\%$ and $\eta_2=R_c/R_2=23.80\%$, where $R_1$ and $R_2$ denote the count rates of channel 1 and channel 2 respectively, and $R_c$ corresponds to the twofold coincidence rate. The overall efficiency is $\eta=R_c/\sqrt{R_1R_2}=23.76\%$.\\
(5) The single photon detector model we used is SPCM-780-10-FC from Excelitas Technologies Corporation. It has a detection efficiency around 67\%, a dark count around 1000 and an after-pulsing probability around 0.3\%. The background count rate is about 1500/s.

\vspace{3mm}
\section*{Appendix B: Quantum process tomography of the $U_a$, $U_b$ and $U_c$ gates}

Before performing QPT on $U_a$, $U_b$, and $U_c$, we first performed QPT on an identity gate (a channel without waveplates), and the fidelity of the reconstructed process with respect to the identity gate is $F = 0.9988\pm0.0005$, which indicates a high level of fidelity of the input state and accuracy of the measurement of the output state.

To implement QPT on $U_a$, $U_b$, $U_c$, six probing states $|H\rangle$, $|V\rangle$, $|D\rangle$, $|A\rangle$, $|R\rangle$ and $|L\rangle$ and three measurement bases X, Y, and Z are used. The measured data are then processed using the maximum likelihood method \cite{Qiang} to calculate the process matrices of $U_a$, $U_b$, and $U_c$. The results are shown in Fig. 4, from which the fidelity of $U_a$, $U_b$, and $U_c$ can be calculated as $0.9902\pm0.0026$, $0.9891\pm0.0027$, $0.9872\pm0.0029$.

\begin{figure}[!htp]
  \centering
  \includegraphics[width=0.45\textwidth]{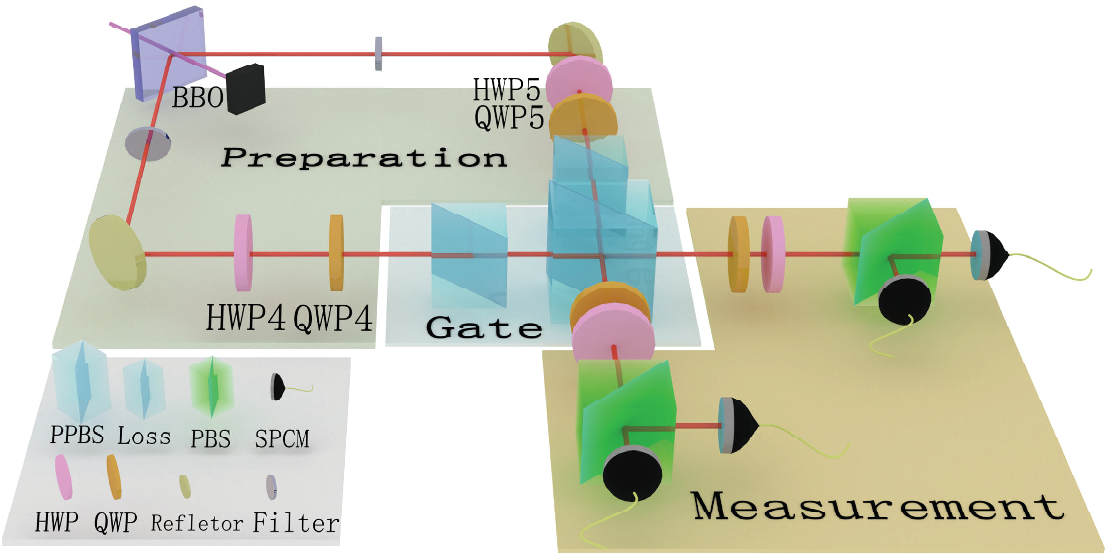}
  \caption{Experimental setup for verifying the CNOT gate. A continuous ultraviolet (UV) laser is focused on a type-II BBO crystal and produces a photon pair. Photon 1 and Photon 2 are prepared at the desired state by tuning HWP4, QWP4, HWP5 and QWP5. After passing through the CNOT gate, which is constructed from three partial polarization beamsplitters (PPBS), photon 1 and photon 2 are then measured separately at the desired basis.}\label{}
\end{figure}

\section*{Appendix C: Quantum gate verification for the two-qubit CNOT gate}
\noindent The experimental setup to implement the two-qubit CNOT gate verification is shown in Fig. 5. The same photon source as in Fig. 2 is used to obtain the two-photon state $|H\rangle_1|V\rangle_2$, which is then prepared to the desired two-qubit state via HWP4, QWP4, HWP5 and QWP5 as the input state for verifying the CNOT gate. One of the following eight two-qubit quantum states
\begin{eqnarray}
\begin{split}
\phi_{IZ}^+&=\frac{I_1}{2}\otimes|0\rangle_2\langle0|,\quad \phi_{IZ}^-=\frac{I_1}{2}\otimes|1\rangle_2\langle1|, \\
\phi_{ZI}^+&=|0\rangle_1\langle0|\otimes\frac{I_2}{2}, \quad\phi_{ZI}^- =|1\rangle_1\langle1|\otimes\frac{I_2}{2},\\
\phi_{IX}^+&= |+\rangle_1\langle+|\otimes\frac{I_2}{2}, \ \ \ \phi_{IX}^-=|-\rangle_1\langle-|\otimes\frac{I_2}{2},\\
 \phi_{XI}^+&=\frac{I_1}{2}\otimes|+\rangle_2\langle+|, \ \ \ \phi_{XI}^-=\frac{I_1}{2}\otimes|-\rangle_2\langle-|
\end{split}
\end{eqnarray}
are prepared randomly with equal probability and then pass through the CNOT gate, which is constructed from three PPBSs \cite{35Okamoto}.

In our experiments, since the pump light is a narrow bandwidth cw laser, the o-light and e-light are anti-correlated in wavelength. Their joint spectral intensity is symmetric with respect to the diagonal and can meet the requirements needed for Hong-Ou-Mandel interference on the PPBS \cite{Hodelin}. To evaluate the performance of the PPBS, we let two H photons interfere on the PPBS and obtained a visibility of $V_{exp}=0.76\pm 0.04$ (the ideal visibility is $V_{th}=0.8$ \cite{36Qiao,Kiesel}).

After passing through the CNOT gate composed of the PPBS, photon 1 and photon 2 are then measured separately at the desired measurement basis. If the initial state is $\phi_{IZ}^+$ ($\phi_{ZI}^+$, $\phi_{IX}^+$, $\phi_{XI}^+$), then it is measured at $ZZ$ ($ZI$, $IX$, $XX$) basis and the verification passes with outcome +1; if the initial state is $\phi_{IZ}^-$ ($\phi_{ZI}^-$, $\phi_{IX}^-$, $\phi_{XI}^-$), then it is measured at $ZZ$ ($ZI$, $IX$, $XX$) basis and the verification passes with outcome -1.
Note that the choice of the measurement basis here is strictly in accordance with the theory of the Ref. \cite{33Liu}, which requires measurements only in the X and Z bases and uses the least number of measurement bases.

The experimental results are shown in Fig. 6, which shows that the infidelity $\epsilon$ decreases with the increase in the number of samples N for both QPT and QGV methods. We note that, since performing a full QPT on a two-qubit gate is time consuming, we only measured one complete set of QPT data with $N\approx324 000$. By sampling this set of data, we obtained a series of QPT data with different numbers of samples. The fidelity of our implemented two-qubit CNOT gate is about $0.8817\pm0.0023$, which is comparable to the fidelity of the reported bulk optical two-qubit entangling gates \cite{36Qiao,37Li}. Although QGV does not show an advantage over QPT in terms of the number of samples in this CNOT experiment, QGV still shows a great advantage in terms of the number of experimental settings with only 16 experimental settings required for QGV, compared to 324 experimental settings for QPT.
\begin{figure}[!htp]
  \flushleft
  \includegraphics[width=0.47\textwidth]{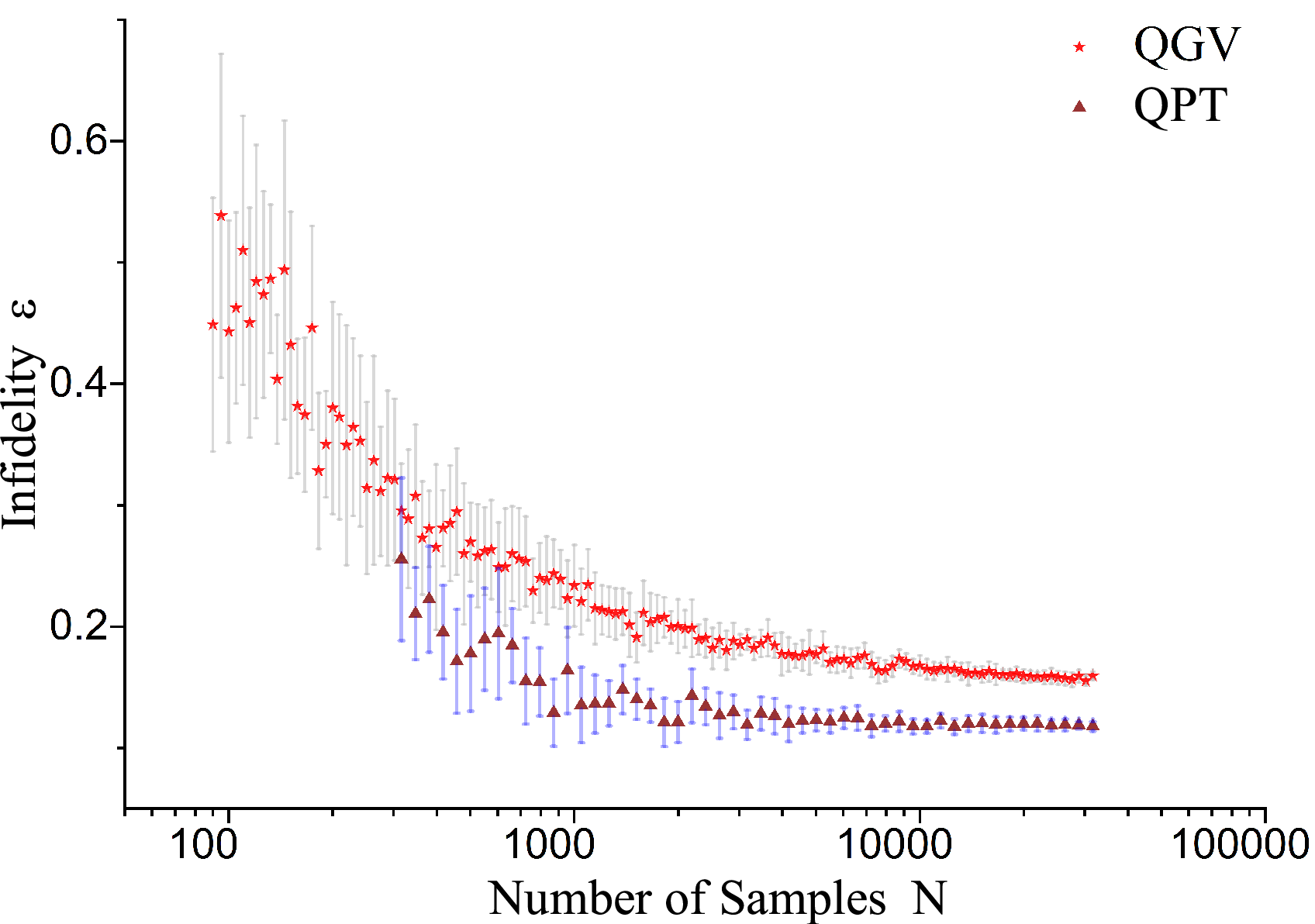}
  \caption{Experimental results of quantum gate verification for the two-qubit CNOT gate. The red five-pointed stars correspond to QGV and the brown triangles correspond to QPT. Here the confidence level $1-\delta$ is set to $99\%$. The error bar of QGV is calculated by repeating the QGV experiment 10 times, whereas the error bar of QPT is directly calculated by assuming a Poissonian statistics for the count rates.}\label{}
\end{figure}


%

\begin{thebibliography}{10}
\bibitem{38}
One hundred arbitrary single-qubit gates are randomly generated using the function RandomUnitary, and for each single-qubit gate, we generate 50 sets of QPT data with sampling numbers $N=10^3, 10^4, 10^5, 10^6, 10^7$ respectively. The average value of the infidelity at each number of samples are then calculated to fit the value of scaling for these 100 quantum gates. These 100 scaling values are all distributed around -0.5, with 3 in the interval [-0.53,-0.52], 24 in the interval [-0.52,-0.51], 31 in the interval [-0.51,-0.5], 27 in the interval [-0.5,-0.49], 9 in the interval [-0.49,-0.48], 5 in the interval [-0.48,-0.47], and 1 in the interval [-0.48,-0.47].
\end{thebibliography}

\end{document}